\input harvmac

\def\K3{{\bf K3}}
\def\journal#1&#2(#3){\unskip, \sl #1\ \bf #2 \rm(19#3) }
\def\andjournal#1&#2(#3){\sl #1~\bf #2 \rm (19#3) }

\def\frac#1#2{{#1\over#2}}

\def\half{\frac12}

\def\inbar{\,\vrule height1.5ex width.4pt depth0pt}
\def\IC{\relax\hbox{$\inbar\kern-.3em{\rm C}$}}
\def\IR{\relax{\rm I\kern-.18em R}}
\def\IP{\relax{\rm I\kern-.18em P}}

%
%


%
\catcode`\@=11
\def\slash#1{\mathord{\mathpalette\c@ncel{#1}}}
\overfullrule=0pt

\def\underrel#1\over#2{\mathrel{\mathop{\kern\z@#1}\limits_{#2}}}

\catcode`\@=12


%


\def \ov {\over}

\def \ha {{1 \ov 2}}

\def \apr {\alpha'}

\def\IL{\relax{\rm I\kern-.18em L}}
\def\IH{\relax{\rm I\kern-.18em H}}
\def\IR{\relax{\rm I\kern-.18em R}}
\def\IC{\relax\hbox{$\inbar\kern-.3em{\rm C}$}}





\def\makeblankbox#1#2{\hbox{\lower\dp0\vbox{\hidehrule{#1}{#2}%
   \kern -#1
   \hbox to \wd0{\hidevrule{#1}{#2}%
      \raise\ht0\vbox to #1{}
      \lower\dp0\vtop to #1{}
      \hfil\hidevrule{#2}{#1}}%
   \kern-#1\hidehrule{#2}{#1}}}%
}%
\def\hidehrule#1#2{\kern-#1\hrule height#1 depth#2 \kern-#2}%
\def\hidevrule#1#2{\kern-#1{\dimen0=#1\advance\dimen0 by #2\vrule
    width\dimen0}\kern-#2}%
\def\openbox{\ht0=1.2mm \dp0=1.2mm \wd0=2.4mm  \raise 2.75pt
\makeblankbox {.25pt} {.25pt}  }

\def\bun#1/#2{\leavevmode
   \kern.1em \raise .5ex \hbox{\the\scriptfont0 #1}%
   \kern-.1em $/$%
   \kern-.15em \lower .25ex \hbox{\the\scriptfont0 #2}%
}

\def\opensquare{\ht0=3.4mm \dp0=3.4mm \wd0=6.8mm  \raise 2.7pt
\makeblankbox {.25pt} {.25pt}  }


\def\sector#1#2{\ {\scriptstyle #1}\hskip 1mm
\mathop{\opensquare}\limits_{\lower
1mm\hbox{$\scriptstyle#2$}}\hskip 1mm}

\def\tsector#1#2{\ {\scriptstyle #1}\hskip 1mm
\mathop{\opensquare}\limits_{\lower
1mm\hbox{$\scriptstyle#2$}}^\sim\hskip 1mm}


\lref\polchinski{J. Polchinski, ``Superstring Theory'', Vol. 1,
Cambridge University Press, Cambridge, 1998.}

\lref\KoganNN{ I.~I.~Kogan and N.~B.~Reis, ``H-branes and chiral
strings,'' Int.\ J.\ Mod.\ Phys.\ A {\bf 16}, 4567 (2001)
[arXiv:hep-th/0107163].
}

\lref\LMS{ H.~Liu, G.~Moore and N.~Seiberg, ``Strings in a
time-dependent orbifold,'' arXiv:hep-th/0204168.
}

\lref\LiuKB{ H.~Liu, G.~Moore and N.~Seiberg,
``Strings in time-dependent orbifolds,''
JHEP {\bf 0210}, 031 (2002) [arXiv:hep-th/0206182].
}

\lref\HartleAI{ J.~B.~Hartle and S.~W.~Hawking, ``Wave Function Of
The Universe,'' Phys.\ Rev.\ D {\bf 28}, 2960 (1983).
}

\lref\VilenkinEV{ A.~Vilenkin, ``Quantum cosmology and eternal
inflation,'' arXiv:gr-qc/0204061.
}

\lref\KhouryBZ{ J.~Khoury, B.~A.~Ovrut, N.~Seiberg,
P.~J.~Steinhardt and N.~Turok, ``From big crunch to big bang,''
Phys.\ Rev.\ D {\bf 65}, 086007 (2002) [arXiv:hep-th/0108187].
}

\lref\KhouryWF{ J.~Khoury, B.~A.~Ovrut, P.~J.~Steinhardt and
N.~Turok, ``The ekpyrotic universe: Colliding branes and the
origin of the hot big  bang,'' Phys.\ Rev.\ D {\bf 64}, 123522
(2001) [arXiv:hep-th/0103239].
}

\lref\SteinhardtVW{ P.~J.~Steinhardt and N.~Turok, ``A cyclic
model of the universe,'' arXiv:hep-th/0111030.
}

\lref\DixonJW{ L.~J.~Dixon, J.~A.~Harvey, C.~Vafa and E.~Witten,
``Strings On Orbifolds,'' Nucl.\ Phys.\ B {\bf 261}, 678 (1985).
}

\lref\DixonJC{ L.~J.~Dixon, J.~A.~Harvey, C.~Vafa and E.~Witten,
``Strings On Orbifolds. 2,'' Nucl.\ Phys.\ B {\bf 274}, 285
(1986).
}

\lref\BachasQT{
C.~Bachas and C.~Hull,
``Null brane intersections,''
JHEP {\bf 0212}, 035 (2002)
[arXiv:hep-th/0210269].
}

\lref\HorowitzMW{ G.~T.~Horowitz and J.~Polchinski, ``Instability
of spacelike and null orbifold singularities,'' Phys.\ Rev.\ D
{\bf 66}, 103512 (2002) [arXiv:hep-th/0206228].
}

\lref\LawrenceAJ{ A.~Lawrence, ``On the instability of 3D null
singularities,'' JHEP {\bf 0211}, 019 (2002)
[arXiv:hep-th/0205288].
}

\lref\MartinecXQ{ E.~J.~Martinec and W.~McElgin, ``Exciting AdS
orbifolds,'' JHEP {\bf 0210}, 050 (2002) [arXiv:hep-th/0206175].
}

\lref\GreeneHU{ B.~R.~Greene, D.~R.~Morrison and A.~Strominger,
``Black hole condensation and the unification of string vacua,''
Nucl.\ Phys.\ B {\bf 451}, 109 (1995) [arXiv:hep-th/9504145].
}

\lref\StromingerCZ{ A.~Strominger, ``Massless black holes and
conifolds in string theory,'' Nucl.\ Phys.\ B {\bf 451}, 96 (1995)
[arXiv:hep-th/9504090].
}

\lref\LiuFT{
H.~Liu, G.~Moore and N.~Seiberg,
``Strings in a time-dependent orbifold,''
JHEP {\bf 0206}, 045 (2002)
[arXiv:hep-th/0204168].
}

\lref\hkmm{J.A. Harvey, D. Kutasov, E. Martinec, and G. Moore,
``Localized Tachyons and RG Flows,''  hep-th/0111154 }

\lref\deconstruction{N. Arkami-Hamed, A.G. Cohen, D.B. Kaplan, A.
Karch, and L. Motl, ``Deconstructing $(2,0)$ and Little String
Theories'' hep-th/0110146}

\lref\atickwitten{J. Atick and E. Witten, ``The Hagedorn
transition and the number of degrees of freedom of string
theory,'' Nucl.Phys.B310:291-334,1988 }

\lref\rohm{R. Rohm, ``Spontaneous supersymmetry breaking in
supersymmetric string theories,''  Nucl.Phys.B237:553,1984 }
\lref\finiteall{G. Moore, ``Finite in All Directions,''
hep-th/9305139}

\lref\horsteif{G.T. Horowitz and A.R. Steif, ``Is spacetime
duality violated in time dependent string solutions?'' Phys. Lett.
{\bf 250B}(1990)49} \lref\smithpolchinski{E. Smith and J.
Polchinski, ``Duality survives time dependence,'' Phys. Lett. {\bf
B} (1991) 59}

\lref\HorowitzAP{ G.~T.~Horowitz and A.~R.~Steif, ``Singular
String Solutions With Nonsingular Initial Data,'' Phys.\ Lett.\ B
{\bf 258}, 91 (1991).
}

\lref\BirrellIX{ N.~D.~Birrell and P.~C.~Davies, ``Quantum Fields
In Curved Space,'' {\it  Cambridge, Uk: Univ. Pr. ( 1982) 340p}. }

\lref\VenezianoPZ{ G.~Veneziano, ``String cosmology: The pre-big
bang scenario,'' arXiv:hep-th/0002094.
}

\lref\KhouryBZ{ J.~Khoury, B.~A.~Ovrut, N.~Seiberg,
P.~J.~Steinhardt and N.~Turok, ``From big crunch to big bang,''
arXiv:hep-th/0108187.
}

\lref\BanadosWN{ M.~Banados, C.~Teitelboim and J.~Zanelli, ``The
Black Hole In Three-Dimensional Space-Time,'' Phys.\ Rev.\ Lett.\
{\bf 69}, 1849 (1992) [arXiv:hep-th/9204099].
}
\lref\BanadosGQ{ M.~Banados, M.~Henneaux, C.~Teitelboim and
J.~Zanelli, ``Geometry of the (2+1) black hole,'' Phys.\ Rev.\ D
{\bf 48}, 1506 (1993) [arXiv:gr-qc/9302012].
}
\lref\ShiraishiHF{ K.~Shiraishi and T.~Maki, ``Quantum fluctuation
of stress tensor and black holes in three-dimensions,'' Phys.\
Rev.\ D {\bf 49}, 5286 (1994).
}

\lref\SteifZV{ A.~R.~Steif, ``The Quantum Stress Tensor In The
Three-Dimensional Black Hole,'' Phys.\ Rev.\ D {\bf 49}, 585
(1994) [arXiv:gr-qc/9308032].
}

\lref\LifschytzEB{ G.~Lifschytz and M.~Ortiz, ``Scalar Field
Quantization On The (2+1)-Dimensional Black Hole Background,''
Phys.\ Rev.\ D {\bf 49}, 1929 (1994) [arXiv:gr-qc/9310008].
}
\lref\KimMI{ H.~Kim, J.~S.~Oh and C.~r.~Ahn, ``Quantisation of
conformal fields in AdS(3) black hole spacetime,'' Int.\ J.\ Mod.\
Phys.\ A {\bf 14}, 2431 (1999) [arXiv:hep-th/9708072].
}
\lref\CarlipQV{ S.~Carlip, ``The (2+1)-Dimensional black hole,''
Class.\ Quant.\ Grav.\  {\bf 12}, 2853 (1995)
[arXiv:gr-qc/9506079].
}
\lref\MaldacenaKR{ J.~M.~Maldacena, ``Eternal black holes in
Anti-de-Sitter,'' arXiv:hep-th/0106112.
} \lref\HorowitzAP{ G.~T.~Horowitz and A.~R.~Steif, ``Singular
String Solutions With Nonsingular Initial Data,'' Phys.\ Lett.\ B
{\bf 258}, 91 (1991).
}
\lref\BalasubramanianRY{ V.~Balasubramanian, S.~F.~Hassan,
E.~Keski-Vakkuri and A.~Naqvi, ``A space-time orbifold: A toy
model for a cosmological singularity,'' arXiv:hep-th/0202187.
}
\lref\GutperleAI{ M.~Gutperle and A.~Strominger, ``Spacelike
branes,'' arXiv:hep-th/0202210.
}
\lref\CornalbaFI{ L.~Cornalba and M.~S.~Costa, ``A New
Cosmological Scenario in String Theory,'' arXiv:hep-th/0203031.
}
\lref\NekrasovKF{ N.~A.~Nekrasov, ``Milne universe, tachyons, and
quantum group,'' arXiv:hep-th/0203112.
}

\lref\HiscockJQ{ W.~A.~Hiscock, ``Quantized fields and chronology
protection,'' arXiv:gr-qc/0009061.
}

\lref\HawkingNK{ S.~W.~Hawking, ``The Chronology protection
conjecture,'' Phys.\ Rev.\ D {\bf 46}, 603 (1992).
}

\lref\HawkingPK{ S.~W.~Hawking, ``The Chronology Protection
Conjecture,'' {\it Prepared for 6th Marcel Grossmann Meeting on
General Relativity (MG6), Kyoto, Japan, 23-29 Jun 1991}. }

\lref\BrodskyDE{ S.~J.~Brodsky, H.~C.~Pauli and S.~S.~Pinsky,
``Quantum chromodynamics and other field theories on the light
cone,'' Phys.\ Rept.\  {\bf 301}, 299 (1998)
[arXiv:hep-ph/9705477].
}

\lref\PolchinskiMF{ J.~Polchinski, ``Critical Behavior Of Random
Surfaces In One-Dimension,'' Nucl.\ Phys.\ B {\bf 346}, 253
(1990).
}

\lref\PolchinskiUQ{ J.~Polchinski, ``Classical Limit Of
(1+1)-Dimensional String Theory,'' Nucl.\ Phys.\ B {\bf 362}, 125
(1991).
}

\lref\HorowitzTA{ G.~T.~Horowitz and R.~C.~Myers, ``The value of
singularities,'' Gen.\ Rel.\ Grav.\  {\bf 27}, 915 (1995)
[arXiv:gr-qc/9503062].
}

\lref\GasperiniBN{ M.~Gasperini and G.~Veneziano, ``The pre-big
bang scenario in string cosmology,'' arXiv:hep-th/0207130.
}

\lref\DiFrancescoSS{ P.~Di Francesco and D.~Kutasov, ``Correlation
functions in 2-D string theory,'' Phys.\ Lett.\ B {\bf 261}, 385
(1991).
}

\lref\DiFrancescoUD{ P.~Di Francesco and D.~Kutasov, ``World sheet
and space-time physics in two-dimensional (Super)string theory,''
Nucl.\ Phys.\ B {\bf 375}, 119 (1992) [arXiv:hep-th/9109005].
}

\lref\MooreGB{ G.~W.~Moore and R.~Plesser, ``Classical scattering
in (1+1)-dimensional string theory,'' Phys.\ Rev.\ D {\bf 46},
1730 (1992) [arXiv:hep-th/9203060].
}

\lref\HawkingSW{ S.~W.~Hawking and R.~Penrose, ``The Singularities
Of Gravitational Collapse And Cosmology,'' Proc.\ Roy.\ Soc.\
Lond.\ A {\bf 314}, 529 (1970).
}

\lref\mwaver{JM.~Figueroa-O'Farrill, ``Breaking the M-waves,''
Class.\ Quant.\ Grav.\ {\bf 17}, 2925 (2000)
[arXiv:hep-th/9904124].}

\lref\BerkoozJE{
M.~Berkooz, B.~Craps, D.~Kutasov and G.~Rajesh,
``Comments on cosmological singularities in string theory,''
arXiv:hep-th/0212215.
}

\lref\KrausIV{
P.~Kraus, H.~Ooguri and S.~Shenker,
``Inside the Horizon with AdS/CFT,''
arXiv:hep-th/0212277.
}

\lref\SimonMA{ J.~Simon, ``The geometry of null rotation
identifications,'' arXiv:hep-th/0203201.
}

\lref\fluxbranes{J.~Figueroa-O'Farrill and J.~Simon, ``Generalized
supersymmetric fluxbranes,'' JHEP {\bf 0112}, 011 (2001)
[arXiv:hep-th/0110170].
}

\lref\FabingerKR{
M.~Fabinger and J.~McGreevy,
``On smooth time-dependent orbifolds and null singularities,''
arXiv:hep-th/0206196.
}

\lref\DixonQV{
L.~J.~Dixon, D.~Friedan, E.~J.~Martinec and S.~H.~Shenker,
``The Conformal Field Theory Of Orbifolds,''
Nucl.\ Phys.\ B {\bf 282}, 13 (1987).
}

\lref\hawkingellis{S.W. Hawking and G.F.R. Ellis,
{\it The large scale structure of space-time},
Cambridge Univ. Press, 1973}


\Title{\vbox{\baselineskip12pt \hbox{gr-qc/0301001}
\hbox{
}}}%
{\vbox{\centerline{ The Challenging Cosmic Singularity$^*$}}}

\smallskip
\centerline{Hong Liu, Gregory Moore}
\medskip

\centerline{\it Department of Physics, Rutgers University}
\centerline{\it Piscataway, New Jersey, 08855-0849}

\bigskip

\centerline{and}

\bigskip

\centerline{Nathan Seiberg\footnote{}{$^*$ Talk delivered by N.
Seiberg at the meeting Challenges to the Standard Paradigm:
Fundamental Physics and Cosmology, November 2002}}
\medskip
\centerline{\it School of Natural Sciences} \centerline{\it
Institute for Advanced Study} \centerline{\it Einstein
Drive,Princeton, NJ 08540}

\smallskip

\vglue .3cm

\bigskip
\noindent In this talk we discuss the cosmic singularity. We
motivate the need to correct general relativity in the study of
singularities, and mention the kind of corrections provided by
string theory.  We review how string theory resolves  time-like
singularities with two examples.  Then, a simple toy model with
lightlike singularities is presented, and studied in
classical string theory.  It turns out that classical string
theory cannot resolve these singularities, and therefore better
understanding of the full quantum theory is needed. The
implications of this result for  the Ekpyrotic/Cyclic Model are
discussed.  We end by mentioning the known suggestions for
explaining the cosmological singularity.

\Date{December 31, 2002}




\newsec{Introduction}

For over thirty years the singularity theorems of  Hawking and
Penrose have presented a challenge to fundamental physics
\refs{\HawkingSW,\hawkingellis}. Among   the most dramatic
consequences of these theorems are the implications for the
standard big bang cosmology: If we extrapolate cosmological
evolution back in time, we are driven to an initial singularity
where all   known physical laws break down. The existence of the
big bang singularity raises some of the most challenging questions
in physics: Is there a beginning of time? What happens before the
big bang?  Are such questions even meaningful and susceptible of
scientific enquiry?

A proper understanding of the singularities requires going beyond
general relativity and introducing new laws of physics. One
promising avenue to doing so is string theory. String theory is a
consistent and finite theory of quantum gravity which reduces to
Einstein's general relativity at low energies and long distances.
It is therefore natural to ask whether, and if so, how, a
singularity in general relativity is resolved in string theory.
By ``resolved'' we mean that the theory yields well-defined
finite answers to physical questions. One hopes that string
theory might  lead to a detailed theory of the big bang. This in
turn might  lead to   observable signatures for quantum
gravitational physics and experimental tests of string theory.

\newsec{Resolution of timelike singularities in string theory}

Over the last two decades advances in string theory have provided
us with various mechanisms for understanding the resolution of
time-independent singularities. These provide essential paradigms
for singularity-resolution in string theory and we will therefore
briefly review two central examples of such singularity
resolution before discussing the big-bang type singularities. In
order to appreciate these resolutions we must first recall the
relation of string theory to general relativity.

At low energies and weak string coupling, the classical and
quantum stringy corrections appear as a generally covariant
extension of the Einstein-Hilbert action with higher curvature
interactions. More precisely, the modifications can be understood
as a double expansion in $\apr/l_c^2 $ and $g_s $. $\apr$ is the
inverse of the string tension, and $l_c$ is a typical curvature
radius.    The string coupling constant plays the role of $\hbar$
and also serves as  the loop counting parameter in string
perturbation theory. For example, with four non-compact
dimensions, the low energy effective action of string theory in
the absence of a cosmological constant has the form
 \eqn\lefa{\eqalign{
 S_{eff} & = {1 \ov g_s^2 \apr} \int d^4 x \sqrt{-g}
 \left(R + a_2 \apr R^2 + \cdots \right) \cr
 & \quad + {1 \ov \apr} \int d^4
 x \sqrt{-g} \left(b_1 R + b_2 \apr R^2 + \cdots \right) \cr
 & \quad + \cdots \cdots
 }}
with the four-dimensional Planck constant  identified as
\eqn\plank{
M_{pl}^2 = {1 \ov g_s^2 \apr} \left(1+ b_1 g_s^2 + \cdots
\right) \ .
}
The first line in \lefa\ corresponds to the classical
contribution.  The second line is the first order quantum
correction.  The expansion in $\apr R$ can be considered as due to
the extended nature of the string; the $g_s = \hbar$ corrections
arise from quantum effects. In particular, the higher order terms
in the first line in \lefa\ correspond to the classical string
modifications of the Einstein gravity. There is no arbitrary UV
cutoff in \lefa. Rather, string theory provides a natural ``UV
cutoff'' in the string scale; this is one of the keys to the UV
finiteness properties of the theory, resolving the long-standing
puzzle of nonrenormalizability of Einstein gravity.

At very low energies or low curvature, and weak string coupling,
the first term in \lefa\ dominates and  one recovers the Einstein
action. However, at high energy or around regions of spacetime
with  large curvature, in which
\eqn\extremes{
 E^2 \apr \sim O(1), \qquad {\rm or} \qquad {E^2 \ov M_{pl}^2}
 \sim O(1)
}
all higher order terms in \lefa\ become important. The low energy
expansion is no longer valid and an exact stringy treatment is
needed. In the two  examples below the expansion \lefa\ breaks
down. Nevertheless, the string theory is susceptible to an exact
treatment, and a satisfying consequence of this treatment is an
understanding of singularity resolution.

In both examples below the   key to singularity resolution is
that, due to the presence of the  spacetime singularity, new
degrees of freedom in string theory become important. These new
degrees of freedom can be classified as {\it perturbative} or
{\it non-perturbative} degrees of freedom, according to how the
masses depend on the string coupling constant $g_s$. The masses
of  perturbative string states have a smooth limit when we take
$g_s \to 0$. These states can be understood as oscillation and
winding modes of fundamental strings. The non-perturbative states
have masses varying as an   inverse power of the string coupling.
Therefore, in smooth backgrounds and at weak coupling $g_s \ll
1$, the non-perturbative degrees of freedom are rather heavy,
they do not contribute in perturbation theory, and their effects
are in general small and relatively unimportant.   By contrast,
around spacetime singularities where the   expansion \lefa\
breaks down   such non-perturbative states actually become
important. Examples of such non-perturbative states include
5-branes,  D-branes and  black holes.

Let  us now turn to our two central examples. In the first
example the singularity is resolved at the classical level, while
in the second example a full non-perturbative treatment is
required.

\medskip

\noindent {\it Example I: Orbifold singularity}

Consider the two dimensional space obtained by identifying
\eqn\orbone{
(x_1 , x_2) \qquad \to \qquad (-x_1, -x_2)
}
in Euclidean   $\IR^2$. The resulting space is a two dimensional
cone with a deficit angle $\pi$. Classical general relativity is
singular in this background because of the delta function
curvature at the tip of the cone. Quantum field theory is also
singular in this background. Moreover, because of the singular
curvature at the orbifold fixed point the expansion \lefa\ breaks
down. Nevertheless, string theory is solvable in this background,
and it turns out that the extended nature of the strings leads to
new degrees of freedom, known as ``twisted states,'' which are
localized near the singularity. These degrees of freedom make the
classical as well as the quantum physics completely smooth
\refs{\DixonJW,\DixonJC}.  Roughly speaking, since the string is
extended, the singularity at the tip of the cone becomes fuzzy
and regular. This is an example of a singularity of classical
general relativity and QFT which is resolved by classical string
effects.

\medskip

\noindent {\it Example II: Conifold singularity}

Let us now consider another example of an orbifold singularity,
the so-called $A_1$ singularity, obtained by taking a $Z_2$
quotient of Euclidean $\IR^4$
\eqn\orbtwo{
(x_1 , x_2, x_3, x_4) \qquad \to \qquad (-x_1, -x_2, -x_3,
 -x_4)  \ .
 }
As a Riemannian manifold, the space is singular at the origin of
$\IR^4$, the fixed point of the identification. In classical
general relativity the singularity at $x^\mu=0$ is   a bolt
singularity and can be ``blown up'' to a spacetime with  a
noncontractible two-sphere, $S^2$, of radius $R$. The orbifold
geometry is then recovered as $R \to 0$. General relativity and
QFT are well-defined for $R>0$, but become singular as $R \to 0$.
It turns out that, again,  the first quantized perturbative
string theory is smooth in this background, even when $R=0$, and
again the underlying reason is the extended nature of the string.
In this case, in addition to the twisted states there is another
crucial ingredient in the resolution. In string theory a
``geometry'' or ``background'' is not simply specified by a
Riemannian metric. There are other quantities which must be
specified, the most important in the present example being the
value of an harmonic 2-form commongly denoted $B_{\mu\nu}$. The
integral
\eqn\beefield{ B= \int_{S^2} \half B_{\mu\nu} dx^\mu \wedge dx^\nu
}
is independent of $R$, and  is a crucial part of the data
specifying a stringy background.  When $R=0$ and $B \not=0$ the
classical string theory is nonsingular. Thus far, the story is
identical to Example I. However, when $R, B \to 0$ there are
further divergences and {\it  classical } string theory turns out
to be singular. Nevertheless, in this case, it turns out that
{\it quantum } string theory is smooth. The essential point is
that one must take into account new {\it non-perturbative}
degrees of freedom which become massless as $R , B \to 0$
\refs{\StromingerCZ,\GreeneHU}.

To summarize, string theory introduces new degrees of freedom. By
including them the physics at timelike singularities becomes regular.
These singularities arise in general relativity simply because the
relevant degrees of freedom have not been taken into account.

\newsec{Cosmological singularities in string theory}

Let us now turn to spacelike and lightlike singularities in
string theory. Such cosmological singularities are much harder to
understand, since the singularity appears in the past or in the
future. Therefore, study of these singularities requires
understanding time-dependent backgrounds - an aspect of string
theory that has been somewhat neglected until recently.

To understand the cosmological singularity in string theory, we
again have to face the question: Is the cosmological singularity
resolved by classical string theory or by quantum string theory?
In this talk we only address the first part of the question,
i.e.\ we study the issue in  string perturbation theory.

As a toy model, we consider a spacetime which is obtained from the
$(2+1)$ flat Minkowski spacetime $\IR^{1,2}$
\eqn\minkmet{
ds^2=-2dx^+dx^-+dx^2
}
by discrete identifications \HorowitzAP\
 \eqn\idenc{\eqalign{
 & x^+ \sim x^+ \cr
 & x \sim x + 2 \pi n x^+ \cr
 & x^- \sim x^- + 2 \pi n x + \ha (2 \pi n)^2 x^+ \cr
 }}
with $n = \pm 1, \pm 2, \cdots$. These identifications are
generated by a Lorentz transformation which is a linear
combination of a boost and a rotation.

To understand better  the geometry of the resulting spacetime, it
is convenient to use the coordinates
 \eqn\newcoordii{ \eqalign{
 x^+ &= y^+ \cr
 x & = y^+ y \cr
 x^- & = y^-  + \ha y^+y^2 \cr} }
in terms of which the metric takes on a gravitational plane wave form:
 \eqn\ymetric{ds^2=-2dy^+dy^-+(y^+)^2dy^2}
subject to the simple identiication   $y \sim y +  2\pi n$. In
terms of these coordinates it is manifest  that the quotient
spacetime is made out of two cones. One of them has $y^+=x^+$
positive and the other has $y^+=x^+$ negative. The radial
direction of the cone $x^+=y^+$ is a null coordinate in the full
spacetime.  The two cones touch at $y^+=x^+=0$ where the space is
singular.  Note that the singularity is neither spacelike nor
timelike; it is lightlike. \foot{We are skating over several
important technicalities in this description. For example, the
coordinate transformation \newcoordii\ is ill defined at $x^+=0$.
Indeed, it turns out that in $X$-coordinates the quotient is
non-Hausdorff at $x^+=0$.}

The above spacetime has a few attractive features:

\item{1.} In \ymetric\ a circle of an infinite size at $y^+ = -
\infty$ shrinks to zero size at $y^+ =0$, and then expands to an
infinite size again at $y^+ = \infty$. This provides an
interesting toy model for understanding the big crunch/big bang
singularity in cosmology.

\item{2.} Time-dependent orbifolds similar to \idenc\ have been
used recently in some proposed cosmological scenarios
\refs{\KhouryWF,\KhouryBZ,\SteinhardtVW}.

\item{3.} As a quotient of flat Minkowski spacetime, the orbifold
\ymetric\ is amenable to exact perturbative string analysis since
we can follow the standard procedure of
\refs{\DixonJW,\DixonJC,\DixonQV}. Furthermore, this orbifold has
some nice properties which make the stringy analysis particularly
clean: it is supersymmetric, there are no closed time-like curves
and there is no particle production.

\item{4.} The same type of singularity also appears in certain
black holes (a closely related problem) as well as in  more
complicated cosmological backgrounds.

Since timelike orbifold singularities are neatly resolved by
classical string theory, as in Example I above, it is natural to
ask if spacelike or lightlike orbifold singularities are likewise
resolved by classical string theory. We should note at the outset
that there are important qualitative differences between the
singularity of Example I and the cosmological singularity at $x^+
= 0$ of \idenc.  For example, in the latter case, the point
$X^\mu=0$, is fixed by an {\it infinite} number of group elements.
Thus, analogies between timelike and lightlike (or spacelike)
orbifolds should be drawn with care.

In order to explore whether it is possible to pass through the
singularity at $x^+ =0$, i.e. whether one can evolve smoothly
from a big crunch to a big bang we will first assume that one can
indeed pass through the singularity. That is, we assume that
classical string theory does indeed somehow resolve the lightlike
singularity. In this case we should be able to  compute S-matrix
elements for particles going from the past cone $x^+ < 0$ to the
future cone $x^+ > 0$. More precisely, we will start in the far
past $x^+ \to - \infty$ with some particles (small fluctuations)
and compute the amplitude to find other particles (other small
fluctuations) in the far future $x^+ \to \infty$ using the
standard rules for orbifolding string backgrounds.  If the
singularity is resolved, then one should be able to find a
sensible S-matrix. On the other hand, a singular S-matrix would
imply the breakdown of the formalism. In \refs{\LiuFT,\LiuKB} we
have looked at certain S-matrix elements in detail at leading
order of string perturbation theory. We found that

\item{$\bullet$} For generic kinematics the amplitudes in
classical string theory are finite.

\item{$\bullet$} However, for special kinematics (near forward
scattering) the string amplitudes diverge.

We should note that, in general relativity on the orbifold
\idenc\ the amplitudes are divergent for generic kinematics. The
finiteness of tree level string amplitudes for generic
kinematics   may be attributed to the softness of strings at high
energies. However, the  divergence in the forward scattering
signals the breakdown of perturbation theory. In fact, the
situation is potentially worse - it is entirely possible that at
higher orders of string perturbation theory the amplitudes are
divergent for generic kinematics. Whether or not this is so is,
at present, unknown.

The physical reason for the divergence is easy to understand
\refs{\LiuKB\HorowitzMW\LawrenceAJ-\MartinecXQ}. Since the
background depends on time, energy is not conserved.  In
particular the energy of an incoming particle is blue-shifted to
infinity by the contraction at the singularity.  The infinite
energy of the incoming particles generates an infinitely large
gravitational field and distorts the geometry. The perturbation
series  breaks down as a result of this large backreaction.

The most likely conclusion we should draw from these computations
is that classical string theory need not resolve the
singularities of  time-dependent   orbifolds. These classical
solutions are unstable and we need to understand the full quantum
theory to explain the physics at the singularity.

We say ``most likely'' because a number of open issues have yet
to be settled. For examples, the possible effects of
brehmstrahlung of light twisted sector states have not yet been
carefully studied. Moreover, the orbifold admits a deformation,
known as the ``nullbrane,''  analogous to the classical
deformation to $R>0$ mentioned in Example II above
\refs{\fluxbranes,\SimonMA,\LiuKB,\HorowitzMW,\FabingerKR}.
String scattering in this background is generically nonsingular.
The limit $R \to 0$ has not yet been carefully studied. For an
entr\'ee into the most recent literature, with references to many
other papers related to our work see
\refs{\BachasQT\BerkoozJE-\KrausIV}.

Let us now put the above computation in the perspective of
enquiries into the nature of the big bang. The idea of going from
a big crunch to a big bang through a non-singular bounce has a
long history. In the thirties, Einstein considered the idea that
a non-singular bounce might be achieved through irregularities.
Richard Tolman considered in detail a cyclic universe using a
similar mechanism. The singularity theorems of Hawking and
Penrose \refs{\HawkingSW,\hawkingellis} ruled out this possibility
in general relativity. The recent suggestions \KhouryBZ\ that the
universe passes through the singularity are motivated by the
orbifold construction of string theory. We now see that classical
string theory is also singular in such orbifolds and cannot be
trusted.  Therefore, the rationale for the proposal of \KhouryBZ\
is absent.  This does not necessarily mean that the proposal is
wrong;  only that it is unmotivated.\foot{The nullbrane at small
radius has a nearly big crunch followed by a nearly big bang. It
might be interesting to explore whether such a geometry is a
reasonable substitute for the cyclic universe.}

There are at least three possibilities for the interpretation of the
singularity:

\item{1.} The singularity is a beginning or end of time. In this
case we need to understand the appropriate initial conditions at
the singularity. For some discussion of these issues see
\refs{\HartleAI, \VilenkinEV}.

\item{2.} Time has no beginning or end. Then one needs to
understand how to pass through the singularity (for recent
discussions see e.g. \refs{\KhouryBZ,\GasperiniBN}).

\item{3.} The most likely possibility, it seems to us,is that
in string theory time is a derived concept.

In conclusion, let us elaborate on the third possibility above.
In toroidal compactifications of string theory there is a minimal
distance, thanks to $T$-duality: shrinking radii past the string
scale does not produce a theory at shorter distances. In more
elaborate compactifications (such as Calabi-Yau compactifications)
it turns out that there can be smooth topology-changing
processes, and ``quantum geometry'' can lead to many
counterintuitive types of behavior.  These examples show that, in
string theory,  standard notions of   topology and geometry  are
not fundamental but are rather   emergent concepts in certain
physical regimes (e.g. in regions of large complex and Kahler
structure parameters, in the Calabi-Yau context).  In another
line of development, Matrix theory and the related advent of
noncommutative field-theoretic limits of string theory further
indicate  that the notion of distance and space ceases to make
sense in certain otherwise sensible regimes of the theory. Given
the principle of  relativity it seems quite likely, perhaps even
inevitable,  that similar statements hold for time as well as for
space.  That is, recent discoveries might be viewed as hints
that   evolution in time might  be only an approximate,
phenomenological,  or emergent concept, which is only applicable
in some but not all physical regimes. But what could we mean by
``physics'' without time? The notion of evolution in time lies at
the very   heart of classical mechanics, of classical (string)
field theory and of the   principles of quantum mechanics.
Finding a satisfactory formulation of physics in which time
evolution is truly an emergent concept is a worthy challenge to
those who would challenge the standard paradigms of fundamental
physics.

\bigskip
\centerline{\bf Acknowledgements}

NS thanks the organizers of the meeting Challenges to the Standard
Paradigm: Fundamental Physics and Cosmology for inviting him to
talk there and for organizing a very exciting and stimulating
meeting.  HL and GM were supported in part by DOE grant 
\#DE-FG02-96ER40949 to Rutgers. NS was supported in part by DOE
grant \#DE-FG02-90ER40542 to IAS.

\listrefs

\end